\documentclass[11pt]{article}
\usepackage{jheppub}
\usepackage{amsmath}
\usepackage{physics}
\usepackage{amssymb}
\usepackage{simpler-wick}
\usepackage{hyperref}
\usepackage{bbold}
\usepackage{graphicx}
\usepackage{pdfpages}
\usepackage{float}
\usepackage{subfig}
\usepackage{makecell}
\usepackage{feynmp-auto}
\usepackage[compat=1.1.0]{tikz-feynman}

\newcommand{\Mpl}{M_\text{Pl}}
\newcommand{\bib}[1]{Ref.~\cite{#1}}
\newcommand{\scat}{\mathcal{M}}

\title{Exploring amplitude criteria for weak gravity in electroweak theory}
\author{Tran Quang Loc}
\affiliation{
The Oscar Klein Centre and Department of Physics, Stockholm University, AlbaNova, 106 91 Stockholm, Sweden}

\abstract{Connections between weak gravity conjecture (WGC) bounds and scattering positivity have been extensively studied over the past decade. This work further explores these connections by proposing positivity as a potential amplitude criterion for weak gravity, with the aim of unifying various weak gravity bounds within a single framework. We illustrate this criterion by analyzing two-body elastic scatterings of photons and Higgs bosons in the forward limit of an electroweak (EW)-like theory. This leads to an amplitude-based criterion that extends magnetic WGC-type bounds to include not only the Abelian gauge coupling but also the non-Abelian gauge and the Yukawa coupling.
Furthermore, a version of the species bound naturally emerges within this setup, suggesting that this amplitude criterion may extend to broader Swampland conjectures beyond the WGC.}
\begin{document}
	\maketitle
	\flushbottom
	\section{Introduction}
The weak gravity conjecture (WGC)~\cite{Arkani-Hamed:2006emk} is a conjectured criterion for determining whether a gravitational effective field theory (EFT) is compatible with quantum gravity. By postulating that gravity is the weakest force, various bounds on non-gravitational interactions have been proposed and studied to date (see~\cite{Harlow:2022ich} for review articles). While the WGC has a wide range of phenomenological implications, it is still not fully understood which interactions and particles in realistic models the conjectured bounds should be applicable to. It is therefore desirable to establish a universal criterion for deriving reasonable WGC-type bounds.

Based on this motivation, we would like to highlight and further explore an interesting similarity between WGC bounds and scattering positivity, which has been observed over the past decade~\cite{Cheung_2014,Heidenreich_2016, Andriolo_2018,Cheung_2014_1,Hamada:2018dde,Bellazzini:2019xts,Arkani_Hamed_2022,Abe_2023,Chen:2019qvr}. The scattering positivity bound is an ultraviolet (UV) constraint on low-energy EFTs that follows from fundamental principles such as unitarity and analyticity of scattering amplitudes. Though its application to gravitational theories involves several non-trivial issues~\cite{Caron-Huot:2024tsk, Hamada:2023cyt, deRham:2022gfe, Alberte:2021dnj} (see Sec.~\ref{sec:gravposi} for details), its potential implications for the Swampland program have been actively explored.

A pioneering work in this direction is Ref.~\cite{Cheung_2014}, which studied four-photon scattering in gravitational QED. The one-loop amplitude exhibits the following low-energy behavior:
\begin{align}
\scat(s,t) = -\frac{s^2}{M_{\rm Pl}^2 t}
+ \left[\frac{e^4}{m_e^4} - \frac{e^2}{M_{\rm Pl}^2 m_e^2} + \mathcal{O}(M_{\rm Pl}^{-4})\right] s^2 + \ldots\,,
\end{align}
in the near-forward region $t \to 0$, where $s$ and $t$ are the standard Mandelstam variables. Here, $e$ and $m_e$ denote the electric gauge coupling and the electron mass, respectively, while $M_{\rm Pl}$ is the reduced Planck mass. The ellipsis indicates higher-order terms, and we have suppressed the $\mathcal{O}(1)$ positive numerical coefficients for simplicity. An interesting observation was that the positivity of the $s^2$ coefficient is ensured as long as the weak gravity bound $e \gtrsim m_e / M_{\rm Pl}$ is satisfied~\cite{Cheung_2014}.

As is well known (and briefly discussed in Sec.~\ref{sec:gravposi}), unlike in non-gravitational theories, the positivity of the $s^2$ coefficient in gravitational theories does not follow solely from the consistency of scattering amplitudes. Therefore, it should not be interpreted as a universal UV constraint on low-energy EFTs. Nevertheless, the observation in Ref.~\cite{Cheung_2014} reveals an intriguing interplay between weak gravity and amplitude behavior. Following this, connections between WGC-type bounds and the sign of low-energy amplitudes have been explored in various contexts, e.g., in the sublattice/tower WGC~\cite{Heidenreich_2016, Andriolo_2018}, the convex-hull WGC~\cite{Chen:2019qvr}, the magnetic WGC~\cite{Cheung_2014_1}, and the black hole WGC~\cite{Hamada:2018dde, Bellazzini:2019xts, Arkani_Hamed_2022, Abe_2023}. Interestingly, studies so far suggest that various versions of the WGC are somewhat related to the positivity of the $s^2$ coefficient and its generalizations that involve UV cutoff scales in low-energy EFTs. Perhaps we may regard the positivity of the $s^2$ coefficient and related inequalities as a candidate for {\it the amplitude criteria for weak gravity} that unifies various weak gravity bounds in a single framework.

One of the advantages of identifying such amplitude criteria is their wide-range applicability. For example, it is known that positivity of the $s^2$ coefficients and related inequalities provides interesting benchmark points in dark sector physics~\cite{Noumi:2022zht,Aoki:2023khq,Kim:2024iud} (for earlier discussions along this line, see e.g.,~\cite{Andriolo_2018,Alberte:2020jsk,Alberte:2021dnj}).

Given this phenomenological importance, we further study connections between WGC bounds and scattering positivity, exploring the amplitude criteria for weak gravity. 
More specifically, we focus on two-body elastic scatterings of photons or Higgs in the Weinberg-Salam (electroweak; EW)-like theory, which is a part of the Standard Model and also provides a simple extension of the previous discussion of \cite{Cheung_2014} to non-abelian gauge theory. We find that the inequality implies magnetic WGC-type bounds on gauge and Yukawa couplings. We also discuss species-type bounds in this context.

The paper is organized as follows: in section~\ref{sec:gravposi}, we review a positivity sum rule for $B^{(2)}(\Lambda)$ in gravitational theories. In section~\ref{sec:bound}, we discuss implications of the inequality for the 2 to 2 scattering of photon and Higgs. Thus, the obtained results are compared with the known WGC in section~\ref{sec:compare}. We conclude in section~\ref{sec:concl}. Several technical details are collected in the appendices.
\section{Motivation: weak gravity versus amplitudes}\label{sec:gravposi}
 \begin{figure}[h]
		\centering
\includegraphics[width=.67\textwidth, trim=60 50 90 40,clip]
        {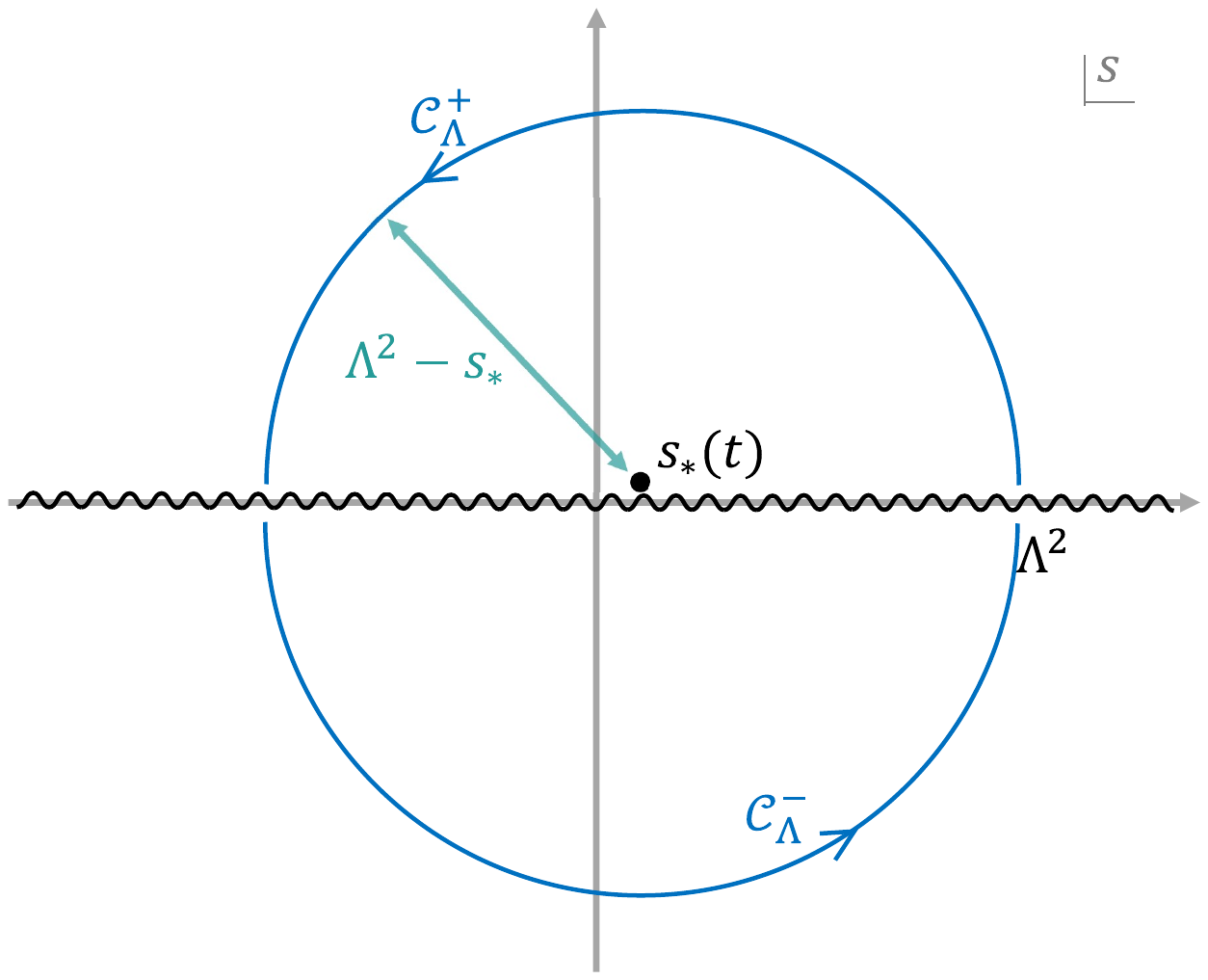}
		\caption{Integration contours used to define $B^{(2)}(\Lambda,t)$ in \eqref{Bdef}. These are semicircles centered at the $s\text{-}u$ crossing symmetric point $s=s_*(t):=(\sum_\text{Ext.}m_\text{ph}^2-t)/2$ with a radius $\Lambda^2-s_*$. We take $s_*$ infinitesimally above the cuts along the real $s$-axis.
        }
		\label{Analytic contour}
	\end{figure}
In this study, we consider the EW model coupled to General Relativity (GR). We interpret this model as an EFT which is valid up to a certain energy scale $\Lambda$. We assume that the EW theory is appropriately embedded into some unknown quantum gravity above the cutoff scale $\Lambda$. 

Within this setup, we consider three distinct 2-by-2 scattering processes involving photons and Higgs bosons: $\gamma \gamma \rightarrow \gamma \gamma$, $H\gamma \rightarrow H\gamma$, and $HH \rightarrow HH$. In this section, however, we simply refer to the corresponding scattering amplitude as $\mathcal{M}(s, t)$. We introduce the usual Mandelstam variables $(s, t, u)$ which satisfy the constraint $s + t + u = \sum_\text{Ext.}m_\text{ph}^2$. Here, the sum is taken over the four external particles, and $m_\text{ph}$ denotes their pole mass, e.g., $m_\text{ph}=m_H$ for the Higgs particle and $m_\text{ph}=0$ for the photon. 

We evaluate the amplitude $\scat(s,t)$ concretely using the EW theory coupled to gravity in the low energy domain, i.e., $|s|,|t|,|u| \leq\Lambda^2$. On the other hand, we postulate several general properties of $\scat$ beyond this low-energy domain without specifying details, and most of them can be summarized into the following twice-subtracted sum rule
\begin{align}
    B^{(2)}(\Lambda, t)
    :=
    &\frac{8}{M_{\mathrm{Pl}}^2 t}+\int_{\mathcal{C}^\pm_\Lambda}\,\frac{\mathrm{d}s}{2\pi i} \frac{\scat(s,t)}{(s-s_*(t))^3}
    \,,\label{Bdef}
\end{align}
where the integral along the contours $\mathcal{C}^\pm_\Lambda$ corresponds to upper and lower arcs centered at the $s\text{-}u$ crossing symmetric point $s_*(t):=(\sum_\text{Ext.}m_\text{ph}^2-t)/2$ with a radius $\Lambda^2-s_*$ in the complex $s$-plane (see Figure~\ref{Analytic contour}).
Here we subtracted the graviton $t$-pole so that $B^{(2)}(\Lambda, t)$ becomes regular in the forward scattering limit $t\to 0$. Thanks to the $s^2$-boundedness and the hermitian analyticity of $\scat(s,t)$, after the contour deformation we obtain the following sum rule for $B^{(2)}(\Lambda,t)$
\begin{align}
     B^{(2)}(\Lambda, t)
    =
    &\frac{8}{M_{\mathrm{Pl}}^2 t}+\frac{4}{\pi} \int_{\Lambda^2}^{\infty} \mathrm{d} s \frac{\operatorname{Im} \mathcal{M}\left(s, t\right)}{\left(s-s_*(t)\right)^3}\,.\label{Br2}
\end{align}
In the absence of gravity, the forward limit of \eqref{Br2} together with unitarity gives the positivity of $B^{(2)}(\Lambda):=B^{(2)}(\Lambda,0)$~\cite{Pham:1985cr,Adams:2006sv}.  In the presence of gravity, however, the forward limit of \eqref{Br2} is nontrivial due to the graviton $t$-pole on the right-hand side (RHS).
This term in the forward limit depends on the details of quantum gravity. However, due to the lack of knowledge of the latter,
there is currently no proof of an inequality $B^{(2)}(\Lambda
)\geq0$ in contrast to the case for non-gravitational theories: a current consensus is that the violation of this inequality of the amount $\mathcal{O}(\Mpl^{-2}m_e^{-2})$ is consistent with the twice-substracted dispersion relation (see e.g., ~\cite{Caron-Huot:2024tsk, Hamada:2023cyt, deRham:2022gfe, Alberte:2021dnj, Caron-Huot:2021rmr, Noumi:2021uuv}).
\begin{figure}[t]
		\centering         \includegraphics[width=4cm]
        {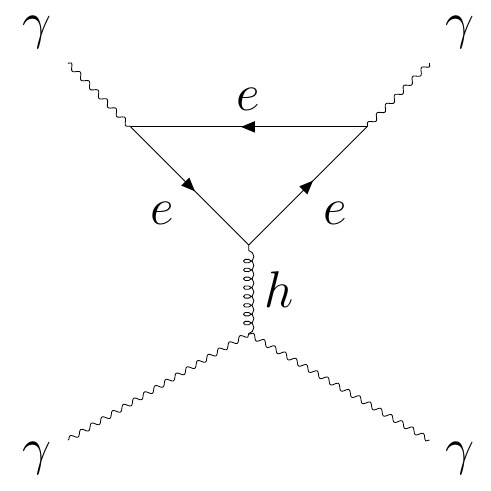}
		\caption{One-loop electron diagram for $\gamma\gamma \to \gamma\gamma$ scattering with graviton $h$ exchange.}
		\label{gagaGRe}
\end{figure}

Intriguingly, the condition $B^{(2)}(\Lambda)\geq 0$ is associated with various forms of the WGC introduced earlier~\cite{Cheung_2014,Heidenreich_2016, Andriolo_2018,Cheung_2014_1,Hamada:2018dde,Bellazzini:2019xts,Arkani_Hamed_2022,Abe_2023,Chen:2019qvr}. Though its extension to a gravitational setup is known to be nontrivial, as we explain above, establishing a similar inequality could offer a useful tool for deriving quantum gravity constraints on EFT, i.e., the Swampland conditions. While this connection is quite suggestive, it has so far been explored only in limited settings, mainly involving (possibly multiple) $U(1)$ gauge theories coupled to gravity. Therefore, our objective is to investigate analogous phenomena within a more comprehensive context, particularly in the EW setup. Such constraints may reveal new insights into the interplay between positivity bounds and quantum gravity criteria, which could be useful for further studies of both in the future.
\section{Bounds on gauge and Yukawa couplings}\label{sec:bound}
In this section, we discuss implications of the bound $B^{(2)}(\Lambda) \geq 0$ for gauge and Yukawa couplings of the EW theory. We first summarize an outline of the EFT calculation of $B^{(2)}(\Lambda)$ and then present the bound $B^{(2)}(\Lambda) \geq 0$ for each scattering process.
\subsection{Outline of EFT calculation}\label{sec:calculation}
We consider $2$-by-$2$ scattering $X_1Y_2 \to X_3Y_4$ in the EW theory coupled to GR, with the external particles $X, Y$ being either the photon or the Higgs boson. For scattering processes involving external photons $\gamma$, we take an $s$–$u$ symmetric sum of helicity amplitudes. For $\gamma\gamma \to \gamma\gamma$, we use
\begin{equation}
	\mathcal{M}=  \frac{1}{4}\left[\mathcal{M}\left(1^{+} 2^{+} 3^{+} 4^{+}\right)+\mathcal{M}\left(1^{+} 2^{-} 3^{+} 4^{-}\right) 
	 +\mathcal{M}\left(1^{-} 2^{-} 3^{-} 4^{-}\right)+\mathcal{M}\left(1^{-} 2^{+} 3^{-} 4^{+}\right)
  \right].\label{gagahelicities}
\end{equation}
Similarly, for $H\gamma\to H\gamma$, we use
\begin{equation}
	\mathcal{M}=  \frac{1}{2}\left[\mathcal{M}\left(1^{H} 2^{+} 3^{H} 4^{+}\right)+\mathcal{M}\left(1^{H} 2^{-} 3^{H} 4^{-}\right) 
\right],\label{gagahelicities}
\end{equation}
where the superscripts $\pm,H$ indicate the type of external particles, i.e., helicity $\pm$ photon and Higgs, respectively. The polarization conventions are specified in Appendix~\ref{Notation Appendix}. We decompose the EFT amplitudes into two parts $\scat_\text{EFT}=\scat_\text{EW}+\scat_\text{GR}$, where $\scat_\text{EW}$ represents the non-gravitational contributions, (i.e., amplitudes in the EW theory without gravity), and $\scat_\text{GR}$ corresponds to processes involving gravity. Correspondingly, we  decompose $B^{(2)}(\Lambda)$ as 
\begin{align}
    B^{(2)}(\Lambda) = 
    B^{(2)}_\text{EW}(\Lambda)
    +
    B^{(2)}_\text{GR}(\Lambda)
    \,.\label{decompose}
\end{align}
We adopt one-loop approximations to evaluate the amplitudes throughout the paper.

The amplitude $\scat_\text{EW}$ should satisfy the twice-subtracted dispersion relation since they consist only of renormalizable vertices. Hence we can use it for evaluating $B^{(2)}_\text{EW}(\Lambda)$, yielding 
\begin{align}
    B^{(2)}_\text{EW}(\Lambda)
    =
    \frac{4}{\pi}\int^\infty_{\Lambda^2}\mathrm{d}s\,
    \frac{\text{Im}\,\scat_\text{EW}(s,0)}{(s-s_*(t))^3}
    \,.\label{eq:EW_sumrule}
\end{align}
This formula shows that for evaluation of $B^\text{(2)}_\text{EW}(\Lambda)$, it is enough to compute diagrams that are dominant in the high-energy amplitude $\text{Im}\,\scat_\text{EW}(s\geq \Lambda^2,0)$. Before presenting quantitative results, it is instructive to estimate the size of $B^{(2)}_{\rm EW}$: The sum rule \eqref{eq:EW_sumrule} can be rewritten as\footnote{
In this estimation we assume $\Lambda^2\gg s_*$ for simplicity.}
\begin{align}
    B^{(2)}_\text{EW}(\Lambda)
    \sim 
    \int^\infty_{\Lambda^2}\,\frac{\sigma_\text{tot}^\text{EW}(s)}{s^2}
    \mathrm{d}s
    \,,
\end{align}
where $\sigma_\text{tot}^\text{EW}(s)$ denotes the total cross section evaluated within the EW sector without gravity. This shows that $B^{(2)}_\text{EW}(\Lambda)$ is positive and calculable from the total cross-section. As we will see below, contributions from massive spin-$1$ particles such as $W,Z$-bosons are dominant at the one-loop level for the processes studied in this paper, which give a constant cross-section $\sigma_\text{tot}^\text{EW}(s\gtrsim m^2)\sim m_{W,Z}^{-2}$ at high energy. This means $B^{(2)}_\text{EW}(\Lambda)\sim m_{W,Z}^{-2}\Lambda^{-2}$. We also have an overall suppression factor by coupling strength, which depends on scattering processes. 
\begin{figure}[t]
 	\centering
 	\includegraphics[width=7cm]{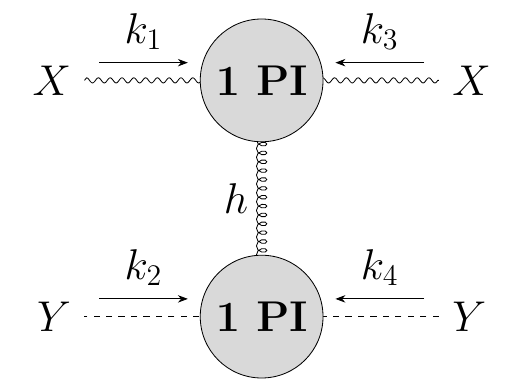}
 	\caption{Diagrams with $t$-channel graviton exchange.}
 	\label{1PI}
 \end{figure}  
For the gravitational part $B^{(2)}_\text{GR}$, the loop diagrams give primary contributions, since the graviton poles are eliminated from its definition. 
We focus on $\mathcal{O}(\Mpl^{-2})$ diagrams in which a single graviton propagator is present. The dominant contribution to $B^{(2)}_{\text{GR}}$ comes from loop corrections to matter-matter-graviton vertices (denoted as $XXh, YYh$) through the $t$-channel graviton exchange diagram represented by Fig.~\ref{1PI}.\footnote{In the $HH \rightarrow HH$ process, we face infrared (IR) divergence due to the graviton's masslessness. This can be addressed by introducing a mass regulator and assuming IR divergence is counteracted by a soft graviton cloud, as suggested in \bib{Noumi:2021uuv}. Under this assumption, it turns out that the contribution from the IR divergent diagrams is of the order of $\Lambda^{-2} M_{\mathrm{Pl}}^{-2}$, which can be discarded in comparison with the $t$-channel exchange.} Formally, this amplitude takes the following form:
\begin{align}
    \mathcal M_\text{fig.\ref{1PI}}(s,t) 
    =
    \frac{R_{XXh}(t)R_{YYh}(t)(-s^2)}{\Mpl^2 t} + \mathcal{O}(s)
    \,,
\end{align}
where $R_{XXh}(t)$ and $R_{YYh}(t)$ characterize the one-particle irreducible (1PI) $XXh$ and $YYh$ vertices, respectively. The gravitational contribution to $B^{(2)}$ is then 
\begin{equation}
    B_{\mathrm{GR}}^{(2)}
    \sim
    \frac{-\partial_t\left[R_{XXh}(t)R_{YYh}(t)\right]_{t=0}}{M_{\mathrm{pl}}^2} 
    \,,
    \label{BGR}
\end{equation}
where we have $\partial_t\left[R_{XXh}(t)R_{YYh}(t)\right]_{t=0}>0$ for the present models. Hence, $B_{\mathrm{GR}}^{(2)}$ is negative, making the inequality $B^{(2)}=B^{(2)}_{\rm EW} +B^{(2)}_{\rm GR}>0$ non-trivial. The $t$-dependence of $R_{XXh}(t)$ and $R_{YYh}(t)$ will be determined by the effective size of the virtual object contributing to the 1PI vertex. For instance, let us consider the $X=\gamma$ case and the electron one-loop corrections to the $\gamma\gamma h$ vertex. In this case, the effective size of the 1PI vertex will be given by the Compton wavelength of an electron-positron pair. We then expect $R'_{XXh}(0)\sim \mathcal{O}(e^2m_e^{-2})$, where a factor $e^2$ is multiplied to account for the dimensionless photon-electron coupling strength. 
\subsection{Bound for each process \label{Bounds}}
It is straightforward to compute both non-gravitational and gravitational contributions to $B^{(2)}$. The results depend on five parameters of the EW theory in addition to the Planck scale \( M_{\text{Pl}} \) and the cutoff scale \( \Lambda \): the Higgs field vacuum expectation value \( v \), the Higgs self-interaction $\lambda$, the $U(1)_{\rm Y}$ and $SU(2)$ gauge couplings \( g_1 \) and \( g_2 \), respectively, and the electron Yukawa coupling \( y_e \). These parameters are related to the particle masses by $m_H=\sqrt{2\lambda}v$ for the Higgs, \( m_W = \frac{v}{2}g_2 \) for the $W$ boson, \( m_Z = \frac{v}{2}\sqrt{g_1^2 + g_2^2} \) for the $Z$ boson, and $m_e = v y_e/\sqrt{2}$ for the electron, respectively. Also recall that the $U(1)_{\rm EM}$ coupling is given by $e=g_1g_2/\sqrt{g_1^2+g_2^2}$. The cutoff scale is supposed to satisfy $\Lambda > v$. Since we are in the weakly coupled regime, the masses of all the particles are sufficiently smaller than the cutoff.
\begin{table}[ht!]
    \centering
    \renewcommand{\arraystretch}{1.7}
    \begin{tabular}{|c|c|c|c|c|}
        \hline
        &\multicolumn{2}{|c|}{\textbf{${HH\rightarrow HH}$ }}&\multicolumn{2}{|c|}{\textbf{${\gamma\gamma\rightarrow \gamma\gamma}$ }}  \\
        \cline{2-5} 
        & $B^{(2)}_\text{EW}$ & $B^{(2)}_\text{GR}$ & $B^{(2)}_\text{EW}$ & $B^{(2)}_\text{GR}$ \\
        \hline
        Z & $\displaystyle\frac{g_1^2+g_2^2}{4\pi^2\Lambda^2 v^2}$ & $\displaystyle\frac{-7}{40 \pi^2 M_{\mathrm{Pl}}^2 v^2}$&nonexistent&nonexistent \\
        W & $\displaystyle\frac{g_2^2}{2\pi^2\Lambda^2 v^2}$ & $\displaystyle\frac{-7}{20 \pi^2 M_{\mathrm{Pl}}^2 v^2}$&$\displaystyle
 \frac{8 e^4}{\pi ^2 \Lambda ^2 v^2 g_2^2 }$&$\displaystyle\frac{-7 e^2}{5 \pi ^2 M_\text{Pl}^2 v^2 g_2^2}$\\
        H &\makecell{$\displaystyle\frac{9 \lambda^2}{4 \pi^2 \Lambda^4}$\\(subleading)} 
        &$\displaystyle -\frac{45-8 \sqrt{3} \pi}{144 \pi^2 M_{\mathrm{Pl}}^2 v^2}$&nonexistent&nonexistent \\
        e & \makecell{$\displaystyle \frac{y_e^4\left(4 \log \left(\Lambda/m_e\right)-3\right)}{ \pi^2 \Lambda^4 }$
        \\(subleading)} & $\displaystyle\frac{-11}{360 \pi^2 M_{\mathrm{Pl}}^2 v^2}$& \makecell{$\displaystyle \frac{e^4\left(4 \log \left(\Lambda/m_e\right)-1\right)}{4\pi^2 \Lambda^4 }$\\(subleading)}
        &$\displaystyle\frac{-11 e^2}{180 \pi ^2 M_\text{Pl}^2 v^2 y_e^2 }$ \\
        \hline
    \end{tabular}\newline
\vspace*{0.1 cm}\newline
    \renewcommand{\arraystretch}{1.7}
    \begin{tabular}{|c|c|c|}
        \hline
        &\multicolumn{2}{|c|}{
        $H\gamma\to H\gamma$
        } \\
        \cline{2-3} 
        & $B^{(2)}_\text{EW}$ & $B^{(2)}_\text{GR}$ \\
        \hline
        Z &nonexistent&$\displaystyle\frac{-7}{80 \pi ^2 M_\text{Pl}^2 v^2}$ \\
        W &$\displaystyle\frac{2e^2}{\pi ^2 \Lambda ^2 v^2}$
        & $\displaystyle\frac{-7}{40 \pi ^2 M_\text{Pl}^2 v^2}-\frac{7e^2 }{10 \pi ^2 M_\text{Pl}^2 v^2 g_2^2 }$\\
        H &nonexistent&$\displaystyle -\frac{45-8 \sqrt{3} \pi }{288 \pi ^2 M_\text{Pl}^2 v^2}$ \\
        e &\makecell{$\displaystyle \frac{e^2 y_e^2\left(4 \log (\Lambda/m_e)+1\right)}{2 \pi^2 \Lambda^4 }$\\(subleading)}&$\displaystyle\frac{-11}{180 \pi ^2 M_\text{Pl}^2 v^2}-\frac{11e^2 }{360 \pi ^2 M_\text{Pl}^2 v^2 y_e^2}$\\
        \hline
    \end{tabular}
    \caption{The quantities $B^{(2)}_\text{EW}$ and $B^{(2)}_\text{GR}$ are calculated for three distinct processes within one-loop approximations: $HH\rightarrow HH$, $\gamma\gamma\rightarrow\gamma\gamma$, and $H\gamma\rightarrow H\gamma$, under the assumption of a light Higgs mass limit. The term ``nonexistent" indicates the absence of certain types of loop contributions involving $Z$/$W$/$H$/$e$, while ``(subleading)" denotes that its preceding expressions are suppressed by $\mathcal{O}(\Lambda^{-4})$.}
    \label{B0Higgmass}
\end{table}

The detailed results of the amplitudes for each scattering process are provided in Appendix~\ref{detailderivation}. However, the full expressions are cumbersome and not particularly illuminating in understanding the bounds due to the presence of various mass scales. The results, however, do not change qualitatively as long as the Higgs is sufficiently light so that the Higgs does not decay into other particles. Hence, for simplicity of presentation, we take the light Higgs limit $m_H \ll m_e, m_W, m_Z$ (i.e. $\lambda \to 0$).\footnote{It is worthwhile noting that our setup should be distinguished from the exact massless Higgs case.} 
The results of $B^{(2)}$ are summarized in Table~\ref{B0Higgmass}.
As we have explained, it is not difficult to understand qualitative behaviours of $B^{(2)}_{\rm EW}$ and $B^{(2)}_{\rm GR}$ without referring to loop calculations explicitly. Let us explain each process in order.
\subsubsection{$H H \rightarrow H H$ \label{intuitive4H}}
The dominant contribution to the non-gravitational part is given by $1/(\Lambda^2 m_{W,Z}^2)$ multiplied by the coupling constants to the Higgs. Since these same coupling constants determine the masses of the $W$ and $Z$ bosons, specifically, $m_W = \frac{v}{2}g_2$ and $m_Z = \frac{v}{2}\sqrt{g_1^2 + g_2^2}$, the gauge coupling dependence in the denominators cancels against the overall coupling constants. As a result, we obtain
\begin{align}
    4\pi^2 B^{(2)}_{\text {EW}}(\Lambda)
    &\simeq\frac{g_2^4}{2\Lambda^2m_W^2}+\frac{(g_1^2+g_2^2)^2}{4\Lambda^2m_z^2}
\nonumber
\\
&
    = \frac{2g_2^2}{\Lambda^2 v^2}+\frac{g_1^2+g_2^2}{\Lambda^2 v^2} 
= \frac{g_1^2+3g_2^2}{\Lambda^2 v^2}
\,,
\end{align}
where the first term and the second term of the first line are from the $W$-boson one-loop and $Z$-boson one-loop diagrams, respectively. The numerical factors need to be determined by explicitly computing the loop integrals. The gravitational contribution can be understood similarly. $B_{\rm GR}^{(2)}$ is determined by the effective size of the 1PI vertex of $HHh$ multiplied by the coupling constants. The coupling dependence is cancelled by the same coupling dependence of the mass, giving the universal form for all the relevant particles, i.e., $e$, $H$, $W$, and $Z$,
\begin{align}
4\pi^2 B^{(2)}_{\rm GR}  &\simeq - \sum_{i=e,H,W,Z} \frac{2n_i^H}{\Mpl^2 v^2} 
= - \frac{2n^H}{\Mpl^2 v^2} 
\,.
\end{align}
The positive numerical factors $n_i^H$ depend on particle species and can be read from Table~\ref{B0Higgmass}. The factor 2 is added because we have two identical contributions from the upper and lower 3-point vertices (see Fig.~\ref{1PI}). We also defined $ 2n^H=2\sum_{i=e,H,W,Z} n_i^H = (125-8\sqrt{3}\pi)/36 \simeq 2.26$. Note that the coupling dependence is different between the non-gravitational and gravitational parts; $B^{(2)}_{\rm EW}$ comes from the 4-point $HHHH$ while $B^{(2)}_{\rm GR}$ is essentially determined by the 3-point $HHh$. As a result, $B^{(2)}_{\rm EW}+B^{(2)}_{\rm GR}\geq 0$ yields a lower bound on the gauge couplings:
\begin{align}
g^2_1+3g^2_2\geq 2n^H\frac{\Lambda^2}{M_\text{Pl}^2}
\,.
\label{EstHH}
\end{align}
\subsubsection{$\gamma \gamma \rightarrow \gamma \gamma$} 
Next, we consider the photon-photon scattering. We can estimate $B^{(2)}_{\rm EW}$ and $B^{(2)}_{\rm GR}$ by noting that the coupling is given by the $U(1)_{\rm EM}$ coupling $e=g_1g_2/\sqrt{g_1^2+g_2^2}$. The non-gravitational part is dominated by the $W$-boson loops of the form,
\begin{align}
4\pi^2 B^{(2)}_\text{EW} \simeq \frac{8e^4}{m_W^2 \Lambda^2}= \frac{32e^4 }{\Lambda^2v^2g_2^2}
\,.
\end{align}
On the other hand, the gravitational contribution is given by
\begin{align}
    4\pi^2 B^{(2)}_{\rm GR}
    &\simeq 
    -\frac{4n_e^{\gamma} e^2}{\Mpl^2 m_e^2} - \frac{n_W^{\gamma} e^2}{2\Mpl^2 m_W^2}
    \nonumber \\
    &=-2n_e^{\gamma} \frac{e^2}{\Mpl^2 v^2 y_e^2}
    -2n_W^{\gamma} \frac{e^2}{\Mpl^2 v^2g_2^2}\,,
\end{align}
with the positive numerical factors $n_e^{\gamma}$ and $n_W^{\gamma}$. Then, the bound $B^{(2)}_{\rm EW}+B^{(2)}_{\rm GR}\geq 0$ reads
\begin{align}
    16\frac{\Mpl^2}{\Lambda^2} &\geq  \frac{n_e^{\gamma}}{y_e^2}\frac{g_2^2}{e^2}+\frac{n_W^{\gamma}}{e^2}
    \nonumber \\
    &=\frac{n_e^{\gamma}}{y_e^2 \sin^2 \theta_{\mathcal{W}}} + \frac{n_W^{\gamma}}{g_1^2} + \frac{n_W^{\gamma}}{g_2^2}
    \,,
    \label{Estgaga}
\end{align}
where $\theta_{\mathcal{W}}$ denotes the Weinberg angle. This inequality now prohibits the individual vanishing of $g_1,g_2$ and $y_e \sin \theta_{\mathcal{W}} $, rather than the combination $g_1^2+3g_2^2$.
\subsubsection{$H \gamma \rightarrow H \gamma$}
For the $H \gamma \rightarrow H \gamma$ process, the overall coupling of the non-gravitational process is given by $e^2$ and the coupling to the Higgs. After the cancellation of the Higgs coupling as in the $HH \to HH$ process, we find
\begin{align}
    4\pi^2 B^{(2)}_{\rm EW} \simeq \frac{8 e^2}{v^2 \Lambda^2}
    \,,
\end{align}
from the $W$ boson loop.
On the other hand, there are two distinct contributions to $B^{(2)}_\text{GR}$ from the $HHh$ vertex and the $\gamma\gamma h$ vertex. Each contribution has been already computed in $HH\to HH$ and $\gamma \gamma \to \gamma \gamma$. We thus obtain
\begin{align}
4\pi^2 B^{(2)}_{\rm GR} \simeq - n^H \frac{1}{\Mpl^2 v^2} -n_e^{\gamma} \frac{e^2}{\Mpl^2 v^2 y_e^2}
    -n_W^{\gamma} \frac{e^2}{\Mpl^2 v^2g_2^2}
    \,.
\end{align}
As a result, $B^{(2)}_{\rm EW}+B^{(2)}_{\rm GR}\geq 0$ yields
\begin{align}
    8\frac{\Mpl^2}{\Lambda^2} &\geq  \frac{n^H}{e^2}+ \frac{n_e^{\gamma}}{y_e^2}+ \frac{n_W^{\gamma}}{g_2^2}
    \nonumber \\
    &=\frac{n_e^{\gamma}}{y_e^2} + \frac{n^H}{g_1^2} + \frac{n^H + n_W^{\gamma}}{g_2^2}
    \,.
    \label{EstgaH}
\end{align}
\section{Comparison with swampland conjectures}\label{sec:compare}
In this section, we compare the bounds obtained in the previous section with Swampland conjectures, especially with the WGC~\cite{Arkani-Hamed:2006emk} and the species bound~\cite{Veneziano:2001ah,Dvali:2007hz}.
\subsection{Weak gravity conjecture}
As we mentioned earlier, a close connection between the bound $B^{(2)}\geq0$ and (the electric version of) the WGC~\cite{Arkani-Hamed:2006emk} has been observed so far, especially in the context of Abelian gauge theories coupled to charged fermions/scalars~\cite{Cheung_2014, Andriolo_2018, Hamada:2018dde, Bellazzini:2019xts, Arkani_Hamed_2022, Abe_2023}. Our purpose here is to explore further connections in the context of the EW theory, as an illustrative example for non-Abelian gauge theories with spontaneous symmetry breaking (see also our earlier work~\cite{Aoki:2021ckh} on the Standard Model).
For reference, let us summarize the bounds $B^{(2)}\geq0$ for the three processes:
\begin{align}
	H H \rightarrow H H:\;\;\;\;\;\;\;\; &g_1^2+3g_2^2
    >
    \frac{125-8\sqrt{3}\pi}{36}\frac{\Lambda^2}{M^2_\text{Pl}}
    \label{MainHH}\,,
    \\
    \gamma \gamma \rightarrow \gamma \gamma:
    \;\;\;\;\;\;\;\; 
    &
     \frac{7}{40}\left(\frac{1}{g_1^2}+\frac{1}{g_2^2}\right)+
     \frac{11}{1440}\frac{1}{y_e^2 \sin^2 \theta_{\mathcal{W}}}
     <
     \frac{M^2_\text{Pl}}{\Lambda^2}\,,\label{Maingaga}
     \\
    H \gamma \rightarrow H \gamma:
    \;\;\;\;\;\;\;\; 
    &
    \frac{125-8 \sqrt{3} \pi}{576}\frac{1}{g_1^2}+\frac{1633-40\sqrt{3}\pi}{2880}\frac{1}{g_2^2}+ 
    \frac{11}{720}\frac{1}{y_e^2}
    <
    \frac{M^2_\text{Pl}}{\Lambda^2}\,,\label{MaingaH}
    \end{align}
where the explicit numerical factors are recovered. Recall that we have ignored the terms called the ``(subleading)'' terms in Table~\ref{B0Higgmass} by assuming a large $\Lambda$. In addition, the light Higgs mass limit, which is achieved by the small Higgs self-coupling limit $g_1,g_2,y_e \gg \lambda$, has been taken to simplify the presentation.

\begin{figure}[h!]
	\centering
 \includegraphics[width=\textwidth]{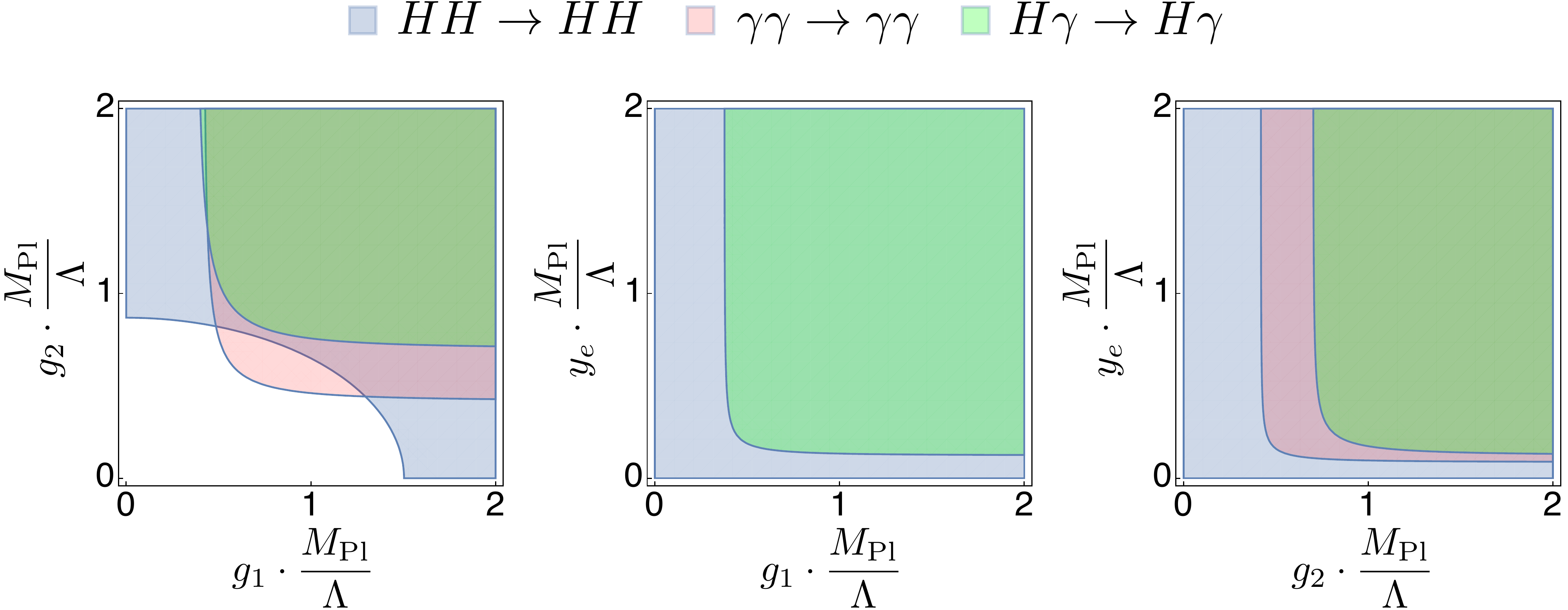}
	\caption{
Projections of constraints on the weak couplings $g_1$, $g_2$, and $y_e$ from gravitational positivity bounds in different limits: (left) $y_e \gg g_1, g_2$; (center) $g_2 \gg g_1, y_e$; (right) $g_1 \gg g_2, y_e$.
}
	\label{CouplingConstraints}
\end{figure}
\begin{figure}[h!]
	\centering
 \includegraphics[width=8cm]{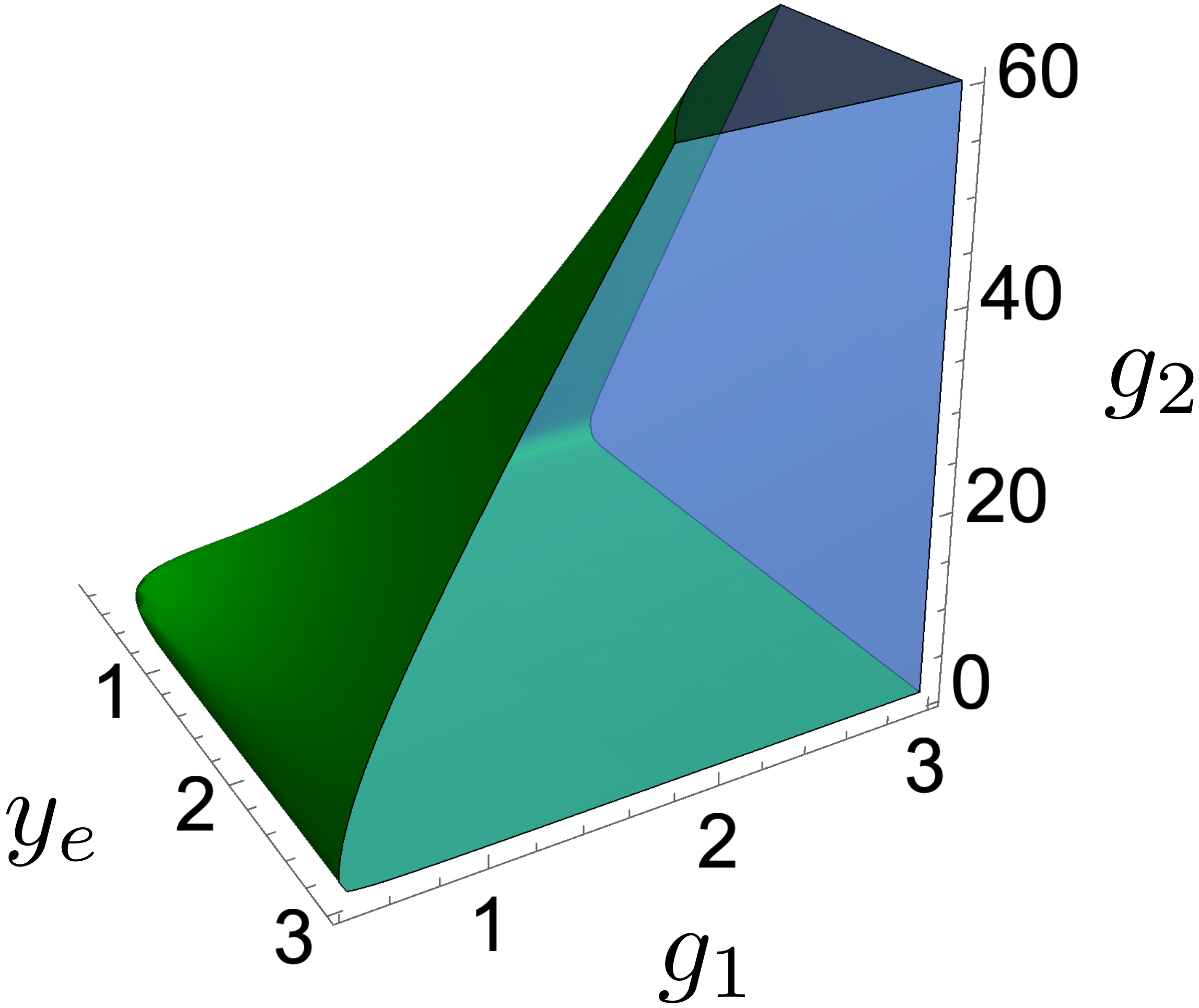}
	\caption{Positivity region of all three processes. Constraint $g_2 \lesssim g_1 y_e \Mpl/\Lambda$ as a consequence of \ref{Maingaga} that carve out the positivity space at large couplings $g_2$. The axes are measured in units of $\Lambda/\Mpl$.}
	\label{g2Inf3d}
\end{figure}
First, we find that all the three bounds \eqref{MainHH}-\eqref{MaingaH} are schematically of the form,
 \begin{align}
    g_1, g_2, y_e > \mathcal{O}(1) \times \frac{\Lambda}{\Mpl}
    \,.
    \label{MWGC-like}
\end{align}
This is similar to the magnetic version of the WGC~\cite{Arkani-Hamed:2006emk}, which gives a bound $\Lambda\lesssim gM_{\rm Pl}$ on the cutoff $\Lambda$ in terms of the gauge coupling $g$ of Abelian gauge theories. A new feature here is that a similar bound is obtained not only for the Abelian gauge coupling $g_1$, but also for the Yukawa coupling $y_e$ (in the same spirit as the scalar WGC~\cite{Palti:2017elp}) and the non-Abelian gauge coupling $g_2$. This demonstrates that various versions of the WGC are packaged into the bound $B^{(2)}\geq0$ in a unified framework. This motivates us to interpret $B^{(2)}\geq0$ as a possible {\it amplitude criterion for weak gravity}.

Second, we observe that various directions in the parameter space of coupling constants can be probed by studying different scattering processes. For example, the $HH\rightarrow HH$ scattering is insensitive to the electron Yukawa coupling $y_e$, but $\gamma\gamma\rightarrow\gamma\gamma$ and $\gamma H\rightarrow \gamma H$ give bounds on $y_e$. Also, it is interesting to notice a qualitative difference between the bound~\eqref{MainHH} and the bounds \eqref{Maingaga}-\eqref{MaingaH}: The bound~\eqref{MainHH} is satisfied even if one of $g_1,g_2$ is zero, as long as the other is large enough. On the other hand, the bounds \eqref{Maingaga}-\eqref{MaingaH} can never be satisfied if any of $g_1,g_2,y_e$ is too small, even if the other couplings are large (it is reminiscent of the convex-hull condition of the WGC for gauge theories with multiple $U(1)$'s~\cite{Cheung:2014vva}). This shows that comprehensive studies of the amplitude
criteria for weak gravity for various scattering processes are useful to efficiently carve out the parameter space of the model.

For illustration, we show the three bounds on two-dimensional subspaces of the parameter space in Fig.~\ref{CouplingConstraints}: The  Fig.~\ref{CouplingConstraints} (left) is a projection of the bounds~\eqref{MainHH}-\eqref{MaingaH} onto the $(g_1,g_2)$-plane under the condition $g_1, g_2\ll y_e\sin\theta_\mathcal{W}$. The allowed region shrinks when we decrease $y_e$ and eventually disappears for $y_e \lesssim \Lambda/\Mpl$. As explained, the constraint from the $HH\rightarrow HH$ scattering requires the gauge couplings to be outside the ellipse $g_1^2+3g_2^2\gtrsim \Lambda^2/\Mpl^2$. The other processes further exclude regions where one of the gauge couplings is individually small. On the other hand, Fig.~\ref{CouplingConstraints} (center) is a projection of the bounds~\eqref{MainHH}-\eqref{MaingaH} onto the $(g_1,y_e)$-plane under the condition $g_1,y_e\ll g_2$. 
Here, the projected region is excluded due to a violation of \eqref{Maingaga}, which imposes an upper bound on $g_2$ relative to $g_1$ and $y_e$, schematically given by $g_2 \lesssim g_1 y_e M_{\rm Pl} / \Lambda$. The corresponding positivity-allowed region in this limit is illustrated in Fig.~\ref{g2Inf3d}. Additionally, the allowed region shrinks when we decrease $g_2$ and eventually disappears for $g_2 \lesssim \Lambda/M_{\rm Pl}$. A similar behavior is observed for $g_1$ in the regime $g_2, y_e \ll g_1$, although in the $(g_2, y_e)$ projection, $g_1$ is not bounded from above by Eq.~\eqref{Maingaga}, in contrast to $g_2$.
 As explained, the $HH\rightarrow HH$ scattering is insensitive to $y_e$, whereas the other bounds carved out the $y_e$-direction.
\subsection{Species bound}
Interestingly, it turns out that the similarity of the amplitude
criteria for weak gravity and swampland conjectures is not limited to the WGC. For illustration, we discuss similarity with the species bound~\cite{Veneziano:2001ah,Dvali:2007hz} by generalizing our analysis to a theory with $N_e$ copies of the electron, while keeping the same gauge symmetry structure, i.e., $U(1)_{\rm Y}\times SU(2)_{\rm L}$ spontaneously broken into $U(1)_{\rm EM}$ by the Higgs mechanism. For simplicity, we assume that all the fermion fields share the same Yukawa coupling $y_e$ and the same charge $e$, and also that there is no large hierarchy among the couplings, i.e., $y_e \sim g_1 \sim g_2$. Our interests are in the large $N_e$ limit with the 't Hooft couplings, defined by $\lambda_e=N_e y_e^2 $ and similarly for gauge couplings, to be fixed in the perturbative regime.

The fermion-loop contributions to $B^{(2)}_{\rm EW}$ and $B^{(2)}_{\rm GR}$ in this setup are obtained by simply multiplying by a factor of $N_e$ in the previous analysis. In the previous analysis, we ignored the fermion contribution to $B^{(2)}_{\rm EW}$ since it was subdominant (see Table~\ref{B0Higgmass}). However, in the present case, the fermion contribution is enhanced by $N_e$, so that it may be non-negligible. Omitting positive numerical factors and logarithm terms $\log(\Lambda/m_e)$, the non-gravitational contributions $B^{(2)}_{\rm EW}$ including the fermion contribution are schematically given by
\begin{align}
	H H \rightarrow H H:\;\;\;\;\;\;\;\; &
 B^{(2)}_{\rm EW}\sim \frac{g_1^2+3g_2^2}{\Lambda^2 v^2} + N_e\frac{y_e^4}{\Lambda^4}
 \,,
    \\
        \gamma \gamma \rightarrow \gamma \gamma:\;\;\;\;\;\;\;\; &
 B^{(2)}_{\rm EW}\sim \frac{e^4}{\Lambda^2 v^2 g_2^2} + N_e \frac{e^4}{\Lambda^4}
 \,,
 \\
	H \gamma \rightarrow H \gamma:\;\;\;\;\;\;\;\; &
 B^{(2)}_{\rm EW}\sim \frac{e^2}{\Lambda^2 v^2} + N_e \frac{e^2 y_e^2}{\Lambda^4}
 \,.
\end{align}
Note that the fermion contribution is negligible if $N_e=1$ and $\Lambda$ is large, which is why it was ignored in the previous analysis. We find that the fermion contribution is dominant if the following conditions are satisfied:
\begin{align}
    \frac{\Lambda^2}{v^2} \lesssim 
    \begin{cases}
        N_e y_e^4/(g_1^2+3g_2^2) & (HH\to HH)\,, \\
        N_e g_2^2 & (\gamma \gamma \to \gamma \gamma) \,, \\
        N_e y_e^2 & (H\gamma \to H \gamma)\,.
    \end{cases}
\end{align}
However, they can never be satisfied as long as the 't Hooft couplings are in the perturbative regime (recall that $\Lambda>v$ for validity of the EFT). We conclude that the fermion contribution to $B^{(2)}_{\rm EW}$ is negligible even in the large $N_e$ limit, at least in the regime of our interests.

Now we can safely recycle our previous results~\eqref{EstHH}, \eqref{Estgaga}, and \eqref{EstgaH} with the fermion contribution to $B^{(2)}_{\rm GR}$ multiplied by a factor of $N_e$. In the large $N_e$ limit, we have
\begin{align}
	H H \rightarrow H H:\;\;\;\;\;\;\;\;
 &g^2_1+3g^2_2 > 2n_e^H \times N_e \frac{\Lambda^2}{M_\text{Pl}^2}
 \,, \label{HHspecies}
 \\
 \gamma \gamma \rightarrow \gamma \gamma:\;\;\;\;\;\;\;\; 
&y_e^2 \sin^2 \theta_{\mathcal{W}}  > \frac{n_e^{\gamma}}{16} \times N_e \frac{\Lambda^2}{\Mpl^2} \,,\label{gagaspecies}
\\
 H \gamma \rightarrow H \gamma:\;\;\;\;\;\;\;\;
 & \frac{n_e^{\gamma}}{8} \frac{N_e}{y_e^2} + \frac{n_e^H}{8} \left(\frac{N_e}{g_1^2} +\frac{N_e}{g_2^2} \right) < \frac{\Mpl^2}{\Lambda^2}\,.
 \label{Hgaspecies}
\end{align}
Schematically, these bounds are of the form
\begin{align}
\label{species_gy_sch}
    g_1, g_2, y_e > \mathcal{O}(1) \times \sqrt{N_e} \frac{\Lambda}{\Mpl}
    \,.
\end{align}
This sharpens the WGC-type bound \eqref{MWGC-like} by a factor $\sqrt{N_e}$. Also, the bound~\eqref{species_gy_sch} is rephrased in terms of the 't Hooft couplings as
\begin{align}
    \lambda_1, \lambda_2, \lambda_e > \mathcal{O}(1) \times N_e \frac{\Lambda}{\Mpl}
    \,,
\end{align}
which has one more factor of $\sqrt{N_e}$. A bound  
\begin{align}
\Lambda\lesssim \frac{M_{\rm Pl}}{\sqrt{N_e}}\,,
\end{align}
analogous to the species bound is obtained if we require $g_1,g_2,y_e\lesssim 1$. However, perturbativity of 't Hooft couplings $\lambda_1,\lambda_2,\lambda_e\lesssim1$ requires an even stronger bound,
\begin{align}
\Lambda\lesssim \frac{M_{\rm Pl}}{N_e}\,.
\end{align}
It would be interesting to study further to understand whether the amplitude
criteria for weak gravity is stronger than the species bound generically, or it is an artifact of the present setup.
\section{Conclusion and discussions}\label{sec:concl}
In our study, we have established a potential link between WGC-like bounds and the principles of positivity within a context of an EW-like theory. Our findings highlight a notable observation: the bounds derived from the interactions of Higgs bosons ($HH \rightarrow HH$) preclude the possibility of both gauge couplings being minimal simultaneously. Moreover, the constraints imposed by the interactions involving photons and Higgs bosons ($\gamma H \rightarrow \gamma H$ and $\gamma\gamma \rightarrow \gamma\gamma$) prohibit any single coupling from being arbitrarily small by itself. This outcome mirrors the implications of the magnetic WGC, indicating that with smaller values of the gauge couplings, $g_1$ and $g_2$, the validity of the effective theory breaks down at a relatively low scale. Furthermore, similar bounds are obtained also for the Yukawa coupling $y_e$, in a spirit akin to the scalar WGC~\cite{Palti:2017elp}. This indicates that multiple incarnations of the WGC are encapsulated within the single bound $B^{(2)} \geq 0$, providing a unified perspective. We are therefore motivated to interpret $B^{(2)} \geq 0$ as a potential \textit{amplitude criterion for weak gravity}. This reinforces the idea that a systematic exploration of amplitude-based constraints across a variety of scattering processes can serve as an efficient and insightful method to delineate the viable parameter space of EFTs under quantum gravity conditions.

On the other hand, by incorporating the scale associated with fermion species, we find that gravitational positivity naturally leads to a version of the species bound. These insights suggest that the interplay between the amplitude
criteria for weak gravity and Swampland conjectures extends beyond the scope of the WGC. In future work, it would be valuable to further explore the implications of these results in more comprehensive Swampland context.
\section{Acknowledgement}
The author would like to
thank Katsuki Aoki, Toshifumi Noumi, and Junsei Tokuda for their collaboration and comments
on the present paper. This work is supported by the ERC Consolidator grant (number: 101125449/acronym: QComplexity). Views and opinions expressed are however those of the authors only and do not necessarily reflect those of the European Union or the European Research Council. Neither the European Union nor the granting authority can be held responsible for them.
\appendix
\section{Polarization Notation \label{Notation Appendix}}
For a scattering $X_1Y_2 \to X_3Y_4$, we consider momenta 1 and 2 as incoming, and momenta 3 and 4 as outgoing. In the center-of-mass (C.O.M.) frame, we adhere to a convention similar to that used in \cite{Alberte:2020bdz}. The momenta and helicities are given by:
\begin{align}
k^\mu_i=(\sqrt{k^2+m_i^2},k \sin \theta_i, 0, k \cos\theta_i)\,,\;\;\;\;\;\;\;
\epsilon^\mu_i(h_i)=\frac{1}{\sqrt{2}}(0,\cos\theta_i,i h_i,-\sin \theta_i)\,.
\end{align}
with massless photons $m_i=0$ and Higgs masses $m_i=m_H$. The convention for the angles $\theta_i$ is as follows 
\begin{align}
\theta_1=0\,,\;\;\;\;\;\;\;\theta_2=\pi\,,\;\;\;\;\;\;\; \theta_3=\theta\,,\;\;\;\;\;\;\; \theta_4=\pi+\theta\,,
\end{align}
where $\theta$ is the scattering angle between $\vec{k_1}$ and $\vec{k_3}$.
$k$ and $\theta$ can be expressed in Mandelstam variables as
\begin{align}
k&=\frac{\sqrt{s}}{2},\;\;\;\;\;\;\;\cos\theta=1+\frac{2t}{s}\,, 
&(\text{for } \gamma\gamma \to \gamma \gamma)\,, \\
k&=\frac{s-m_H^2}{2\sqrt{s}},\;\;\;\;\;\;\;\cos\theta=1+\frac{2ts}{s-m_H^2}\,,
&(\text{for } H\gamma \to H\gamma)\,,
\end{align}
respectively.
The amplitude reads
\begin{align}
\mathcal{M}_{\gamma\gamma \to \gamma\gamma}\left(h_1, h_2, h_3, h_4\right)&=\epsilon_1^\mu\left(h_1\right) \epsilon_2^\nu\left(h_2\right) \mathcal{M}_{\mu \nu \alpha \beta}\left(k_1, k_2, k_3, k_4\right) \epsilon_3^{* \alpha}\left(h_3\right) \epsilon_4^{* \beta}\left(h_4\right), \\
\mathcal{M}_{H\gamma \to H\gamma}(h_1,h_3)&=\epsilon_1^\mu(h_1)\mathcal{M}_{\mu\nu}(k_i)\epsilon_3^{*\nu}(h_3)
\,.
\end{align}
We will omit the subscript if the process we study is obvious from the context.
The useful relations between momenta and helicities are
\begin{equation}
\begin{aligned}
& \epsilon_{12}=\frac{1}{2}(1+h_1h_2)\,, \quad \epsilon_{13}=\frac{1}{2}\left(-1-h_1h_3-\frac{2t}{s}\right)\,, \quad \epsilon_{14}=\frac{1}{2}\left(1-h_1h_4+\frac{2t}{s}\right)\,, \\
& \epsilon_{23}=\frac{1}{2}\left(1-h_2h_3+\frac{2t}{s}\right)\,, \quad \epsilon_{24}=\frac{1}{2}\left(-1-h_2h_4-\frac{2t}{s}\right)\,, \quad \epsilon_{34}=\frac{1}{2}(1+h_3h_4)\,,
\end{aligned}
\end{equation}
\begin{equation}
(k \epsilon)_{13}=(k \epsilon)_{24}=\sqrt{\frac{t u}{2s}}\,, \quad(k \epsilon)_{14}=(k \epsilon)_{23}=-\sqrt{\frac{t u}{2s}}\,.
\end{equation}
for $\gamma\gamma \to \gamma \gamma$ where $(k \epsilon)_{i j} \equiv k_i \cdot \epsilon_j\text{ (or } k_i \cdot \epsilon^*_j)=-k_j \cdot \epsilon_i \text{ (or } -k_j \cdot \epsilon^*_i)$ and $\epsilon_{i j} \equiv \epsilon^*_i \cdot \epsilon_j= \epsilon^*_j \cdot \epsilon_i$. On the other hand, the inner products of the polarization vector and the momenta for $H\gamma \to H\gamma$ are
\begin{align}
k_1\cdot\epsilon_3=-k_3\cdot\epsilon_1=-k_2\cdot\epsilon_3=k_4\cdot\epsilon_1&=\frac{k\sin\theta}{\sqrt{2}}=\frac{\left(s-m_H^2\right) \sqrt{1-\left(\frac{2 s
			t}{\left(m_H^2-s\right)^2}+1\right)^2}}{2 \sqrt{2} \sqrt{s}}\,,\\
\epsilon_1\cdot\epsilon^*_3=\epsilon_3\cdot\epsilon^*_1&=\frac{1}{2} \left(-h_1 h_3-\frac{2 s
	t}{\left(s-m_H^2\right)^2}-1\right)\,.
\end{align}	
\section{Bounds beyond light Higgs mass limit}\label{detailderivation}
As we have explained in Sec.~\ref{sec:calculation}, the non-gravitational part of $B^{(2)}$ is computed by the cross-section while the gravitational part is dominated by the one-loop graviton exchange diagram Fig.~\ref{1PI}. In the main text, we considered a light Higgs mass limit to simplify the expressions. In this appendix, we provide the full expressions without taking such a limit and discuss the bounds for a general Higgs mass. We only discuss $HH\to HH$ and $H\gamma \to H\gamma$ because $\gamma \gamma \to \gamma \gamma$ is not affected by the Higgs at the one-loop order.

Note that the non-gravitational part is computed by the tree-level cross-sections, so the mass of the external line is not significant to the results. Hence, the non-gravitational part $B^{(2)}_{\rm EW}$ given in Table~\ref{B0Higgmass} is applicable even when we keep a finite $m_H$. Furthermore, the leading contribution of the gravitational part is given by the graviton $t$-channel exchange, which is controlled by the 1-loop corrections to the three-point vertices, $HHh$ and $\gamma\gamma h$, as shown in eq.~\eqref{BGR}. Therefore, all we need for a finite Higgs mass is the full expressions for the 1-loop corrections to the $HHh$ vertex. Since the calculation of the 1PI vertex is lengthy but straightforward, we directly show the final result of $B^{(2)}_{\rm GR}$.

\begin{figure}[h!]
 	\centering
 	\includegraphics[width=15cm]{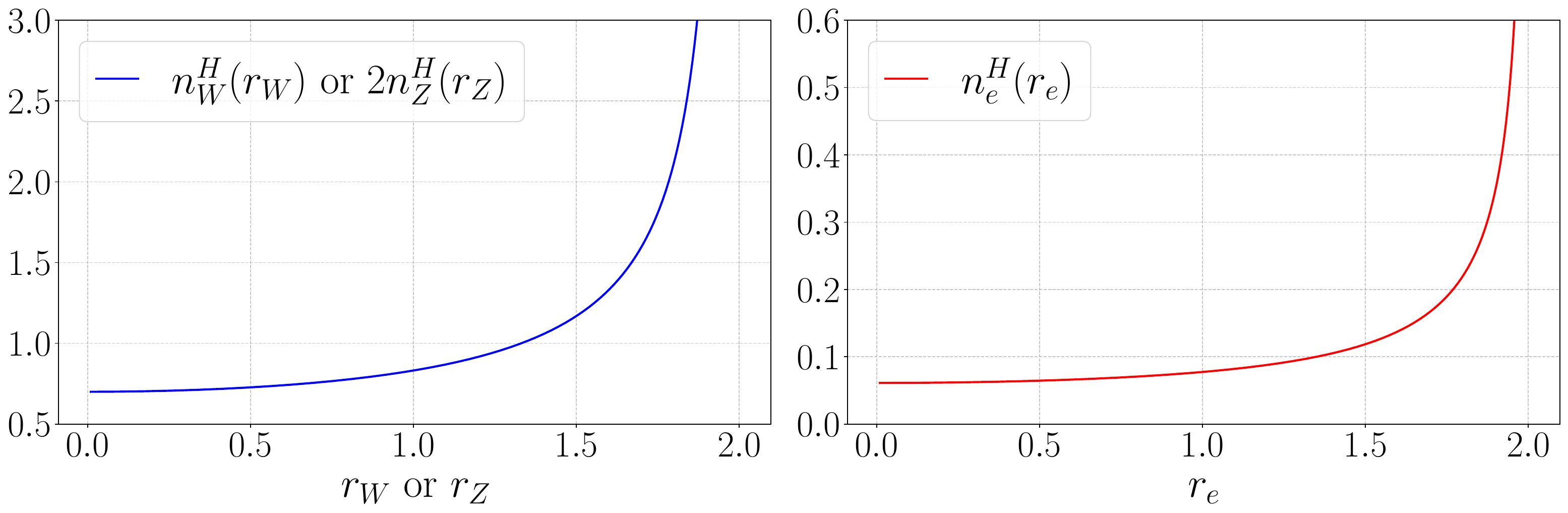}
 	\caption{$n^H_i$ as a function of $r_i$ for $i=W,Z,e$}
 	\label{4H-cGR-Wboson-kW}
 \end{figure}

Let's first take an example of the $HH\to HH$ process. 
We consider positive Higgs mass under the decaying threshold, $0<m_H<2m_\text{loop}$, where $m_\text{loop}$ denotes the mass of the particle that appears in the loop.\footnote{In general, the positivity of the $s^2$ coefficient does not hold for unstable external states even in non-gravitational theories~\cite{Aoki:2022qbf,Aoki:2023tmu}.} We define the ratio of the Higgs mass to the loop mass:
\begin{align}
    r_i=m_H/m_i\,, \quad (i=e, W,Z)
    \,.
\end{align}
which takes $r_i \in (0,2)$. Because of $m_H^2=2v^2\lambda$, varying $r_i$ corresponds to varying $\lambda$ within the range of positive Higgs mass below the decay threshold. The decay threshold restricts the upper bound on the Higgs mass to be smaller than $ \min[2m_e, 2m_Z, 2m_W]$, which imposes an upper limit on $\lambda$ given by $\min\left[2\frac{m_e^2}{v^2}, \frac{g_2^2}{2}, \frac{g_1^2 + g_2^2}{2}\right]$, or equivalently, $\min\left[2\frac{m_e^2}{v^2}, \frac{g_2^2}{2}\right]$. The gravitational part $B^{(2)}_{\rm GR}$ is computed as
\begin{align}
4\pi^2 B^{(2)}_\text{GR} &= -\sum_{i=e,H,W,Z}  \frac{2n_i^H}{M_\text{Pl}^2v^2}
\,,
\end{align}
where
\begin{align}
   n_e^H=&
\frac{12 r_e^2 + r_e^4 - r_e^6 + \sqrt{r_e^2(-4 + r_e^2)} (6 + r_e^2) \log \left(\frac{1}{2} \left(2 - r_e^2 + \sqrt{r_e^2(-4 + r_e^2)}\right)\right)}{3 r_e^6 (-4 + r_e^2)}\,,\label{neH}
    \\n_W^H=&
\frac{1}{12 r_W^6 (-4 + r_W^2)^2} \bigg[ (-2 + r_W) r_W^2 (2 + r_W) (-72 - 60 r_W^2 + 14 r_W^4 + r_W^6) + \nonumber \\&\left.4 \sqrt{r^2 (-4 + r^2)} (36 + 24 r_W^2 - 29 r_W^4 + 5 r_W^6) \log \left(\frac{1}{2} \left(2 - r_W^2 + \sqrt{r_W^2 (-4 + r_W^2)} \right)\right) \right]\,,\label{nZH}
\\
 n_Z^H = &\frac{n_W^H}{2},
\intertext{and}
    n_H^H = &\frac{45-8\sqrt{3}\pi}{72}\,.
\end{align}
     The functions $n_W^H$ and $n_e^H$ range in $\left[\frac{7}{10}, \infty\right)$ and $\left[\frac{11}{180},\infty\right)$  as $r_i$ goes from 0 to $2$ as shown in Fig.~\ref{4H-cGR-Wboson-kW}. The positivity of $B^{(2)}$ gives the bound on the gauge couplings in the same way as \eqref{EstHH} by replacing $n^H_i$ with the $r_i$-dependent quantities \eqref{neH}-\eqref{nZH}. We recover the result in the main text by taking $r_i\to 0$. For a finite Higgs mass, $n_i^H(r_i)$ remains a number of the order of unity as long as the Higgs mass is sufficiently smaller than the decay thresholds. Hence, the Higgs mass does not change our conclusion qualitatively in that mass region.
 On the other hand, $n_i^H(r_i)$ diverges at the decay threshold $r_i=2~(\Leftrightarrow m_H=2m_i)$, which is also the threshold of the triangle singularity in the forward kinematics. The gravitational part then negatively diverges, implying that the gravitational positivity $B^{(2)}_{\text{non-grav}}+B^{(2)}_{\rm GR}>0$ would no longer be satisfied. One possible interpretation of this is that such a mass spectrum is prohibited in quantum gravity. Alternatively, one can think of it as an example that the gravitational positivity does not work as a swampland criterion if this mass spectrum can be consistently realised. It would be interesting to explore the near-decay threshold from a different perspective of quantum gravity.

 The bound from $H\gamma \to H\gamma$ is similarly obtained. The positivity of $B^{(2)}$ reads \eqref{Estgaga} after replacing the numerical factors with \eqref{neH}-\eqref{nZH}. Thus, we do not discuss $H\gamma \to H\gamma$ further.
\nocite{*}
\bibliographystyle{JHEP}
\bibliography{ref}

\providecommand{\href}[2]{#2}\begingroup\raggedright\begin{thebibliography}{10}

\bibitem{Arkani-Hamed:2006emk}
N.~Arkani-Hamed, L.~Motl, A.~Nicolis and C.~Vafa, \emph{{The String landscape, black holes and gravity as the weakest force}}, \href{http://dx.doi.org/10.1088/1126-6708/2007/06/060}{\emph{JHEP} {\bf 06} (2007) 060}, [\href{https://arxiv.org/abs/hep-th/0601001}{{\tt hep-th/0601001}}].

\bibitem{Harlow:2022ich}
D.~Harlow, B.~Heidenreich, M.~Reece and T.~Rudelius, \emph{{Weak gravity conjecture}}, \href{http://dx.doi.org/10.1103/RevModPhys.95.035003}{\emph{Rev. Mod. Phys.} {\bf 95} (2023) 035003}, [\href{https://arxiv.org/abs/2201.08380}{{\tt 2201.08380}}].

\bibitem{Cheung_2014}
C.~Cheung and G.~N. Remmen, \emph{Infrared consistency and the weak gravity conjecture}, \href{http://dx.doi.org/10.1007/jhep12(2014)087}{\emph{Journal of High Energy Physics} {\bf 2014} (Dec., 2014) }.

\bibitem{Heidenreich_2016}
B.~Heidenreich, M.~Reece and T.~Rudelius, \emph{Sharpening the weak gravity conjecture with dimensional reduction}, \href{http://dx.doi.org/10.1007/jhep02(2016)140}{\emph{Journal of High Energy Physics} {\bf 2016} (Feb., 2016) }.

\bibitem{Andriolo_2018}
S.~Andriolo, D.~Junghans, T.~Noumi and G.~Shiu, \emph{A tower weak gravity conjecture from infrared consistency}, \href{http://dx.doi.org/10.1002/prop.201800020}{\emph{Fortschritte der Physik} {\bf 66} (May, 2018) }.

\bibitem{Cheung_2014_1}
C.~Cheung and G.~N. Remmen, \emph{Naturalness and the weak gravity conjecture}, \href{http://dx.doi.org/10.1103/physrevlett.113.051601}{\emph{Physical Review Letters} {\bf 113} (July, 2014) }.

\bibitem{Hamada:2018dde}
Y.~Hamada, T.~Noumi and G.~Shiu, \emph{{Weak Gravity Conjecture from Unitarity and Causality}}, \href{http://dx.doi.org/10.1103/PhysRevLett.123.051601}{\emph{Phys. Rev. Lett.} {\bf 123} (2019) 051601}, [\href{https://arxiv.org/abs/1810.03637}{{\tt 1810.03637}}].

\bibitem{Bellazzini:2019xts}
B.~Bellazzini, M.~Lewandowski and J.~Serra, \emph{{Positivity of Amplitudes, Weak Gravity Conjecture, and Modified Gravity}}, \href{http://dx.doi.org/10.1103/PhysRevLett.123.251103}{\emph{Phys. Rev. Lett.} {\bf 123} (2019) 251103}, [\href{https://arxiv.org/abs/1902.03250}{{\tt 1902.03250}}].

\bibitem{Arkani_Hamed_2022}
N.~Arkani-Hamed, Y.-t. Huang, J.-Y. Liu and G.~N. Remmen, \emph{Causality, unitarity, and the weak gravity conjecture}, \href{http://dx.doi.org/10.1007/jhep03(2022)083}{\emph{Journal of High Energy Physics} {\bf 2022} (Mar., 2022) }.

\bibitem{Abe_2023}
Y.~Abe, T.~Noumi and K.~Yoshimura, \emph{Black hole extremality in nonlinear electrodynamics: a lesson for weak gravity and festina lente bounds}, \href{http://dx.doi.org/10.1007/jhep09(2023)024}{\emph{Journal of High Energy Physics} {\bf 2023} (Sept., 2023) }.

\bibitem{Chen:2019qvr}
W.-M. Chen, Y.-T. Huang, T.~Noumi and C.~Wen, \emph{{Unitarity bounds on charged/neutral state mass ratios}}, \href{http://dx.doi.org/10.1103/PhysRevD.100.025016}{\emph{Phys. Rev. D} {\bf 100} (2019) 025016}, [\href{https://arxiv.org/abs/1901.11480}{{\tt 1901.11480}}].

\bibitem{Caron-Huot:2024tsk}
S.~Caron-Huot and J.~Tokuda, \emph{{String loops and gravitational positivity bounds: imprint of light particles at high energies}},  \href{https://arxiv.org/abs/2406.07606}{{\tt 2406.07606}}.

\bibitem{Hamada:2023cyt}
Y.~Hamada, R.~Kuramochi, G.~J. Loges and S.~Nakajima, \emph{{On (scalar QED) gravitational positivity bounds}}, \href{http://dx.doi.org/10.1007/JHEP05(2023)076}{\emph{JHEP} {\bf 05} (2023) 076}, [\href{https://arxiv.org/abs/2301.01999}{{\tt 2301.01999}}].

\bibitem{deRham:2022gfe}
C.~de~Rham, S.~Jaitly and A.~J. Tolley, \emph{{Constraints on Regge behavior from IR physics}}, \href{http://dx.doi.org/10.1103/PhysRevD.108.046011}{\emph{Phys. Rev. D} {\bf 108} (2023) 046011}, [\href{https://arxiv.org/abs/2212.04975}{{\tt 2212.04975}}].

\bibitem{Alberte:2021dnj}
L.~Alberte, C.~de~Rham, S.~Jaitly and A.~J. Tolley, \emph{{Reverse Bootstrapping: IR Lessons for UV Physics}}, \href{http://dx.doi.org/10.1103/PhysRevLett.128.051602}{\emph{Phys. Rev. Lett.} {\bf 128} (2022) 051602}, [\href{https://arxiv.org/abs/2111.09226}{{\tt 2111.09226}}].

\bibitem{Noumi:2022zht}
T.~Noumi, S.~Sato and J.~Tokuda, \emph{{Phenomenological motivation for gravitational positivity bounds: A case study of dark sector physics}}, \href{http://dx.doi.org/10.1103/PhysRevD.108.056013}{\emph{Phys. Rev. D} {\bf 108} (2023) 056013}, [\href{https://arxiv.org/abs/2205.12835}{{\tt 2205.12835}}].

\bibitem{Aoki:2023khq}
K.~Aoki, T.~Noumi, R.~Saito, S.~Sato, S.~Shirai, J.~Tokuda et~al., \emph{{Gravitational Positivity for Phenomenologists: Dark Gauge Boson in the Swampland}},  \href{https://arxiv.org/abs/2305.10058}{{\tt 2305.10058}}.

\bibitem{Kim:2024iud}
S.~Kim and P.~Ko, \emph{{Constraining Millicharged dark matter with Gravitational positivity bounds}},  \href{https://arxiv.org/abs/2405.04454}{{\tt 2405.04454}}.

\bibitem{Alberte:2020jsk}
L.~Alberte, C.~de~Rham, S.~Jaitly and A.~J. Tolley, \emph{{Positivity Bounds and the Massless Spin-2 Pole}}, \href{http://dx.doi.org/10.1103/PhysRevD.102.125023}{\emph{Phys. Rev. D} {\bf 102} (2020) 125023}, [\href{https://arxiv.org/abs/2007.12667}{{\tt 2007.12667}}].

\bibitem{Pham:1985cr}
T.~N. Pham and T.~N. Truong, \emph{{Evaluation of the Derivative Quartic Terms of the Meson Chiral Lagrangian From Forward Dispersion Relation}}, \href{http://dx.doi.org/10.1103/PhysRevD.31.3027}{\emph{Phys. Rev. D} {\bf 31} (1985) 3027}.

\bibitem{Adams:2006sv}
A.~Adams, N.~Arkani-Hamed, S.~Dubovsky, A.~Nicolis and R.~Rattazzi, \emph{{Causality, analyticity and an IR obstruction to UV completion}}, \href{http://dx.doi.org/10.1088/1126-6708/2006/10/014}{\emph{JHEP} {\bf 10} (2006) 014}, [\href{https://arxiv.org/abs/hep-th/0602178}{{\tt hep-th/0602178}}].

\bibitem{Caron-Huot:2021rmr}
S.~Caron-Huot, D.~Mazac, L.~Rastelli and D.~Simmons-Duffin, \emph{{Sharp boundaries for the swampland}}, \href{http://dx.doi.org/10.1007/JHEP07(2021)110}{\emph{JHEP} {\bf 07} (2021) 110}, [\href{https://arxiv.org/abs/2102.08951}{{\tt 2102.08951}}].

\bibitem{Noumi:2021uuv}
T.~Noumi and J.~Tokuda, \emph{{Gravitational positivity bounds on scalar potentials}}, \href{http://dx.doi.org/10.1103/PhysRevD.104.066022}{\emph{Phys. Rev. D} {\bf 104} (2021) 066022}, [\href{https://arxiv.org/abs/2105.01436}{{\tt 2105.01436}}].

\bibitem{Veneziano:2001ah}
G.~Veneziano, \emph{{Large N bounds on, and compositeness limit of, gauge and gravitational interactions}}, \href{http://dx.doi.org/10.1088/1126-6708/2002/06/051}{\emph{JHEP} {\bf 06} (2002) 051}, [\href{https://arxiv.org/abs/hep-th/0110129}{{\tt hep-th/0110129}}].

\bibitem{Dvali:2007hz}
G.~Dvali, \emph{{Black Holes and Large N Species Solution to the Hierarchy Problem}}, \href{http://dx.doi.org/10.1002/prop.201000009}{\emph{Fortsch. Phys.} {\bf 58} (2010) 528--536}, [\href{https://arxiv.org/abs/0706.2050}{{\tt 0706.2050}}].

\bibitem{Aoki:2021ckh}
K.~Aoki, T.~Q. Loc, T.~Noumi and J.~Tokuda, \emph{{Is the Standard Model in the Swampland? Consistency Requirements from Gravitational Scattering}}, \href{http://dx.doi.org/10.1103/PhysRevLett.127.091602}{\emph{Phys. Rev. Lett.} {\bf 127} (2021) 091602}, [\href{https://arxiv.org/abs/2104.09682}{{\tt 2104.09682}}].

\bibitem{Palti:2017elp}
E.~Palti, \emph{{The Weak Gravity Conjecture and Scalar Fields}}, \href{http://dx.doi.org/10.1007/JHEP08(2017)034}{\emph{JHEP} {\bf 08} (2017) 034}, [\href{https://arxiv.org/abs/1705.04328}{{\tt 1705.04328}}].

\bibitem{Cheung:2014vva}
C.~Cheung and G.~N. Remmen, \emph{{Naturalness and the Weak Gravity Conjecture}}, \href{http://dx.doi.org/10.1103/PhysRevLett.113.051601}{\emph{Phys. Rev. Lett.} {\bf 113} (2014) 051601}, [\href{https://arxiv.org/abs/1402.2287}{{\tt 1402.2287}}].

\bibitem{Alberte:2020bdz}
L.~Alberte, C.~de~Rham, S.~Jaitly and A.~J. Tolley, \emph{{QED positivity bounds}}, \href{http://dx.doi.org/10.1103/PhysRevD.103.125020}{\emph{Phys. Rev. D} {\bf 103} (2021) 125020}, [\href{https://arxiv.org/abs/2012.05798}{{\tt 2012.05798}}].

\bibitem{Aoki:2022qbf}
K.~Aoki, \emph{{Unitarity and unstable-particle scattering amplitudes}}, \href{http://dx.doi.org/10.1103/PhysRevD.107.065017}{\emph{Phys. Rev. D} {\bf 107} (2023) 065017}, [\href{https://arxiv.org/abs/2212.05670}{{\tt 2212.05670}}].

\bibitem{Aoki:2023tmu}
K.~Aoki and Y.-t. Huang, \emph{{Anomalous thresholds for the S-matrix of unstable particles}}, \href{http://dx.doi.org/10.1007/JHEP09(2024)045}{\emph{JHEP} {\bf 09} (2024) 045}, [\href{https://arxiv.org/abs/2312.13520}{{\tt 2312.13520}}].

\bibitem{Herrero-Valea:2022lfd}
M.~Herrero-Valea, A.~S. Koshelev and A.~Tokareva, \emph{{UV graviton scattering and positivity bounds from IR dispersion relations}}, \href{http://dx.doi.org/10.1103/PhysRevD.106.105002}{\emph{Phys. Rev. D} {\bf 106} (2022) 105002}, [\href{https://arxiv.org/abs/2205.13332}{{\tt 2205.13332}}].

\bibitem{Herrero-Valea:2020wxz}
M.~Herrero-Valea, R.~Santos-Garcia and A.~Tokareva, \emph{{Massless positivity in graviton exchange}}, \href{http://dx.doi.org/10.1103/PhysRevD.104.085022}{\emph{Phys. Rev. D} {\bf 104} (2021) 085022}, [\href{https://arxiv.org/abs/2011.11652}{{\tt 2011.11652}}].

\bibitem{Hebecker:2023qwl}
A.~Hebecker, J.~Jaeckel and R.~Kuespert, \emph{{Small kinetic mixing in string theory}}, \href{http://dx.doi.org/10.1007/JHEP04(2024)116}{\emph{JHEP} {\bf 04} (2024) 116}, [\href{https://arxiv.org/abs/2311.10817}{{\tt 2311.10817}}].

\bibitem{Caron-Huot:2022ugt}
S.~Caron-Huot, Y.-Z. Li, J.~Parra-Martinez and D.~Simmons-Duffin, \emph{{Causality constraints on corrections to Einstein gravity}}, \href{http://dx.doi.org/10.1007/JHEP05(2023)122}{\emph{JHEP} {\bf 05} (2023) 122}, [\href{https://arxiv.org/abs/2201.06602}{{\tt 2201.06602}}].

\bibitem{Noumi:2022wwf}
T.~Noumi and J.~Tokuda, \emph{{Finite energy sum rules for gravitational Regge amplitudes}}, \href{http://dx.doi.org/10.1007/JHEP06(2023)032}{\emph{JHEP} {\bf 06} (2023) 032}, [\href{https://arxiv.org/abs/2212.08001}{{\tt 2212.08001}}].

\bibitem{Palti:2020tsy}
E.~Palti, \emph{{Fermions and the Swampland}}, \href{http://dx.doi.org/10.1016/j.physletb.2020.135617}{\emph{Phys. Lett. B} {\bf 808} (2020) 135617}, [\href{https://arxiv.org/abs/2005.08538}{{\tt 2005.08538}}].

\bibitem{Alberte_2021}
L.~Alberte, C.~de~Rham, S.~Jaitly and A.~J. Tolley, \emph{Qed positivity bounds}, \href{http://dx.doi.org/10.1103/physrevd.103.125020}{\emph{Physical Review D} {\bf 103} (June, 2021) }.

\bibitem{Denner:1991kt}
A.~Denner, \emph{{Techniques for calculation of electroweak radiative corrections at the one loop level and results for W physics at LEP-200}}, \href{http://dx.doi.org/10.1002/prop.2190410402}{\emph{Fortsch. Phys.} {\bf 41} (1993) 307--420}, [\href{https://arxiv.org/abs/0709.1075}{{\tt 0709.1075}}].

\bibitem{Bellazzini:2021shn}
B.~Bellazzini, G.~Isabella, M.~Lewandowski and F.~Sgarlata, \emph{{Gravitational causality and the self-stress of photons}}, \href{http://dx.doi.org/10.1007/JHEP05(2022)154}{\emph{JHEP} {\bf 05} (2022) 154}, [\href{https://arxiv.org/abs/2108.05896}{{\tt 2108.05896}}].

\bibitem{Tokuda:2020mlf}
J.~Tokuda, K.~Aoki and S.~Hirano, \emph{{Gravitational positivity bounds}}, \href{http://dx.doi.org/10.1007/JHEP11(2020)054}{\emph{JHEP} {\bf 11} (2020) 054}, [\href{https://arxiv.org/abs/2007.15009}{{\tt 2007.15009}}].

\bibitem{Cheung:2014ega}
C.~Cheung and G.~N. Remmen, \emph{{Infrared Consistency and the Weak Gravity Conjecture}}, \href{http://dx.doi.org/10.1007/JHEP12(2014)087}{\emph{JHEP} {\bf 12} (2014) 087}, [\href{https://arxiv.org/abs/1407.7865}{{\tt 1407.7865}}].

\bibitem{Linden:2024fby}
T.~Linden, T.~T.~Q. Nguyen and T.~M.~P. Tait, \emph{{X-Ray Constraints on Dark Photon Tridents}},  \href{https://arxiv.org/abs/2406.19445}{{\tt 2406.19445}}.

\bibitem{Tran:2023lzv}
V.~Q. Tran, T.~T.~Q. Nguyen and T.-C. Yuan, \emph{{Self-interacting vectorial dark matter in a SM-like dark sector}}, \href{http://dx.doi.org/10.1088/1475-7516/2024/05/015}{\emph{JCAP} {\bf 05} (2024) 015}, [\href{https://arxiv.org/abs/2312.10785}{{\tt 2312.10785}}].

\end{thebibliography}\endgroup
\end{document}